\documentclass[10pt]{article}

\usepackage{amsmath}
\usepackage{amssymb}

\usepackage{graphicx}

\usepackage{cite}

\usepackage{placeins}

\usepackage{color} 

\usepackage[labelfont=bf,labelsep=period,justification=raggedright]{caption}

\makeatletter
\renewcommand{\@biblabel}[1]{\quad#1.}
\makeatother

\date{}

\usepackage{dcolumn}
\usepackage{bm}

\begin{document}

\begin{flushleft}
{\Large
\textbf{Epidemic centrality - is there an underestimated epidemic impact of network
peripheral nodes?}
}
\\
Mile \v{S}iki\'{c}$^{1,5,\ast}$, 
Alen Lan\v{c}i\'{c}$^{2}$,
Nino Antulov-Fantulin$^{3}$,
Hrvoje \v{S}tefan\v{c}i\'{c}$^{4}$
\\
\bf{1} Bioinformatics Institute, A*STAR, Singapore
\\
\bf{2} Faculty of Science, Department of Mathematics, University of Zagreb,
Zagreb, Croatia
\\
\bf{3} Division of Electronics, Rudjer Bo\v{s}kovi\'{c} Institute, Zagreb,
Croatia
\\
\bf{4} Theoretical Physics Division, Rudjer Bo\v{s}kovi\'{c} Institute, Zagreb,
Croatia
\\
\bf{5} Faculty of Electrical Engineering and Computing, University of Zagreb,
Zagreb, Croatia
\\
$\ast$ E-mail: Corresponding mile.sikic@fer.hr

\end{flushleft}

\section*{Abstract}

In the study of disease spreading on empirical complex networks in SIR model, 
initially infected nodes can be ranked according to some measure of their
epidemic impact. The highest ranked nodes, also referred to as ``superspreaders",
are associated to dominant epidemic risks and therefore deserve special
attention. In simulations on studied empirical complex networks, it is shown that
the ranking depends on the dynamical regime of the disease spreading. A possible
mechanism leading to 
this dependence is illustrated in an analytically tractable example.
In 
systems where the allocation of resources to counter disease spreading to
individual nodes is based on their ranking, the dynamical regime of disease
spreading is frequently not known before the outbreak of the disease. Therefore,
we introduce a quantity called {\em epidemic centrality} as an average over all
relevant regimes of disease spreading as a basis of the ranking. A recently introduced
concept of phase diagram of epidemic spreading is used as a framework in which
several types of averaging are studied. The epidemic centrality is compared to
structural properties of nodes such as node degree, k-cores and betweenness. 
There is a growing trend of epidemic centrality with degree and k-cores values,
but the variation of epidemic centrality is much smaller than the variation of
degree or k-cores value. It is found that the epidemic centrality of the
structurally peripheral nodes is of the same order of magnitude as the epidemic
centrality of the structurally central nodes. The implications of these findings
for the distributions of resources to counter disease spreading are discussed.

\section*{Author Summary}

Studies of disease spreading on complex networks have provided a deep insight
into the conditions of onset, dynamics and prevention of epidemics in human
populations and 
malicious software propagation in computer networks. Identifying nodes which,
when initially infected, on average infect the largest part of the network and
ranking 
them according to their epidemic impact (the portion of the network eventually
infected) is a priority for public health policies. In the study of epidemic
spreading on 
empirical complex networks in the Susceptible-Infected-Recovered model, we find
that the required ranking depends on the disease spreading regime, i.e. on how
fast 
the disease is transmitted between nodes and how fast the infected node
recovers. A measure called {\em epidemic centrality}, averaging the epidemic
impact over all 
possible disease spreading regimes, is introduced as a basis of epidemic
ranking.  We find the epidemic centrality of nodes which are structurally
central, to be of the same order of magnitude as the
 epidemic centrality of structurally peripheral nodes. These
findings point to the need to study if the impact of an epidemic starting at structurally
peripheral nodes might be considerably underestimated. Network periphery should gain a more
prominent role in the study of the allocation of resources in future epidemic preparedness
plans. 

\section*{Introduction}


The spreading of contagious diseases represents one of the most dangerous and
disruptive  phenomena in human communities and animal populations
\cite{may,keeling}. 
The propagation of malicious software in computer and communication networks is
a technological counterpart of spreading of contagious diseases
\cite{Vespignani}. 
The pathways of spreading of detrimental disturbances in these systems are well
described by complex networks \cite{rev1,rev2,rev3}. The dynamics of spreading
of 
diseases on complex networks and mathematical models of such spreading in
general \cite{revspread} make a subject of considerable interest and activity of
research 
community and of big practical importance. 

Other forms of spreading are also present in systems described as complex
networks. The dissemination of information, formation of public opinion or
spreading of 
fashion proceed in a very similar way as the spreading of diseases, see e.g.
section 6.1 in \cite{it}. The studies of these  specific forms of social
dynamics have 
attracted a lot of interest of academic community and recently the potential of
their commercial application is increasingly coming into focus. 

The research on empirical complex networks has revealed their very heterogeneous
structure \cite{rev1,rev2,rev3}. In particular, in scale-free networks the nodes
with 
degrees differing many orders of magnitude may coexist. Therefore, it is not
surprising that different nodes have different importance in spreading of
disease or 
information over the network. Finding the nodes that contribute the most to the
spreading or the network structures that are the most robust to disease 
propagation \cite{Youssef} is essential in planning disease control and
prevention or devising efficient 
network marketing strategies. In general, it is important to rank nodes
according to their {\em epidemic impact}. In this paper we refer to such ranking
as {\em epidemic ranking}. The nodes with the highest epidemic impact,
frequently referred to as ``key players'' or ``superspreaders'', are usually 
identified using the structural properties of the underlying complex network.
They have been identified with nodes of high degree (hubs) \cite{dd1,dd2,dd3},
high values of k-cores \cite{kcores} or betweenness centrality
\cite{betw1,betw2}. Based on these network structural properties it is possible
to construct some ranking of nodes regarding their spreading capabilities. It
should be stressed that our definition of superspreading nodes differs somewhat
from the concept of superspreaders used in the literature.
In epidemiological literature \cite{sars1,lloyd,het}, the superspreaders are
defined as infected nodes which produce a large number of secondary cases (for
a 
precise quantitative definition of a superspreading event see \cite{lloyd}). In
this paper we are interested in the type of superspreading at the level of the
entire 
network, i.e. we are interested in initially infected nodes which lead to a
large number of infected nodes in the entire network. In this paper we define
 a ``superspreader'' as a node which when initially infected, leads to a very
large number of infected nodes at the level of the entire network.  To emphasize
this 
difference we put quotation marks around the word superspreader. We find this
generalization from the secondary cases to the entire network natural and
convenient 
for the present paper.

The spreading dynamics does not depend solely on structural properties. The
intensity of spreading of disease is also controlled by properties such as its
transmission 
rate and average infectious period of the infected node. In this paper we adopt
the stochastic Susceptible-Infected-Recovered (SIR) epidemiological model
\cite{sir}. 
The discrete time stochastic dynamics in this model is controlled by two
parameters: $p$, the probability per time step that the infected node infects a
neighboring 
susceptible node, and $q$, the probability per time step that the infected node
recovers. An important question is how the dynamics of spreading affects the
status of 
``superspreaders'' and the epidemic ranking in total. In particular, if the node
A has a higher epidemic ranking than the node B in one dynamical regime of
spreading 
(e.g. for SIR model parameters $(p_1,q_1)$), does the ordering of their rankings
(i.e. epidemic impacts) remain the same in some other dynamical regime of
spreading 
(for some other SIR model parameters $(p_2,q_2)$)?

The principal aim of this paper is to study the dependence of the
epidemic ranking on the dynamical regime of the disease spreading and compare it to the ranking derived from structure. An
important tool in 
achieving this aim is {\em the phase diagram of epidemic spreading}, a
diagrammatic representation of epidemic impact for all possible epidemic
parameter values and for 
a given network and a given initially infected node \cite{miPhysA}. Using the
concept of phase diagram of epidemic spreading, a measure of {\em epidemic
centrality} 
that takes the spreading dynamics into account is proposed. Finally, the
implications to security policies and an optimal allocation of resources are
discussed. 

In our considerations in this paper we shall assume that the network is static,
i.e. that its structure does not change during the course of disease spreading. 
For a complex network of social contacts this assumption is certainly just a
starting approximation since during the outbreak social contacts may in general
vary  
and get restricted \cite{volz,schwartz}. For information and communication
networks or spreading of ideas the assumption of static underlying complex
network is a 
much better approximation. 
 
The paper is organized in the following way. The first section is the
Introduction. The second section of Materials and Methods comprises three
subsections: the 
first subsection focuses on the dependence of the ``superspreader''
ranking of a node on the dynamical regime of the disease spreading. The second subsection summarizes the concept of phase
diagram 
of epidemic spreading. In the third subsection the measure of epidemic
centrality is introduced. The final section of Results and Discussion is devoted
to the 
analysis of results of simulations on empirical complex networks and their
discussion and the paper closes with the summary and conclusions. 


\section*{Materials and Methods}

A majority of results presented in this paper have been obtained by simulations
on empirical complex networks. In particular, the following networks have been
used: complex network of 2003 condensed
matter collaborations (with  27 519 nodes) hereafter referred to as {\em
cond-mat 2003} network \cite{collab}, an undirected,
unweighted network representing the topology of the US Western States Power Grid
 (having 4941 nodes) hereafter referred to as {\em power grid} network
\cite{net1}, network of
coauthorships between scientists posting preprints on the Astrophysics E-Print
Archive between Jan 1, 1995 and December
31, 1999, (consisting of 16 706 nodes) hereafter referred to as {\em astro-ph}
network \cite{collab}, and a symmetrized snapshot of the structure of the
Internet at the level of autonomous
systems, reconstructed from BGP tables posted by the University of Oregon Route
Views Project (containing 22 963
nodes) hereafter referred to as {\em internet} network \cite{net3}. To
complement our research on empirical complex networks, we also perform studies
on 
synthetic Erdos-Renyi (ER) networks. The ER type of networks is chosen to test 
the concepts introduced in this paper also on networks significantly different
from networks with broad degree distributions.

\subsection*{Dependence of the "superspreader" status on the disease spreading dynamical regime}
\label{dyn}

Our principal hypothesis is that the node which is highly ranked as a
``superspreader'' for some disease spreading parameters (i.e. in some disease
spreading regime) may not be highly ranked for some other epidemic parameters
(i.e. in some other epidemic regime). More generally, the epidemic ranking of a
node $i$ according to its epidemic impact, measured by e.g. the average number
of infected nodes for an initially infected node $i$, is dependent on the
dynamics of disease spreading, i.e. parameters describing the spreading of the
disease.  In this section this hypothesis is tested and supported in two ways.
First we present results of simulations on empirical complex networks and then
an analytical calculation for a specific artificial network is displayed.

\begin{figure}[!ht]
\begin{center}
\includegraphics[width=0.5\textwidth]{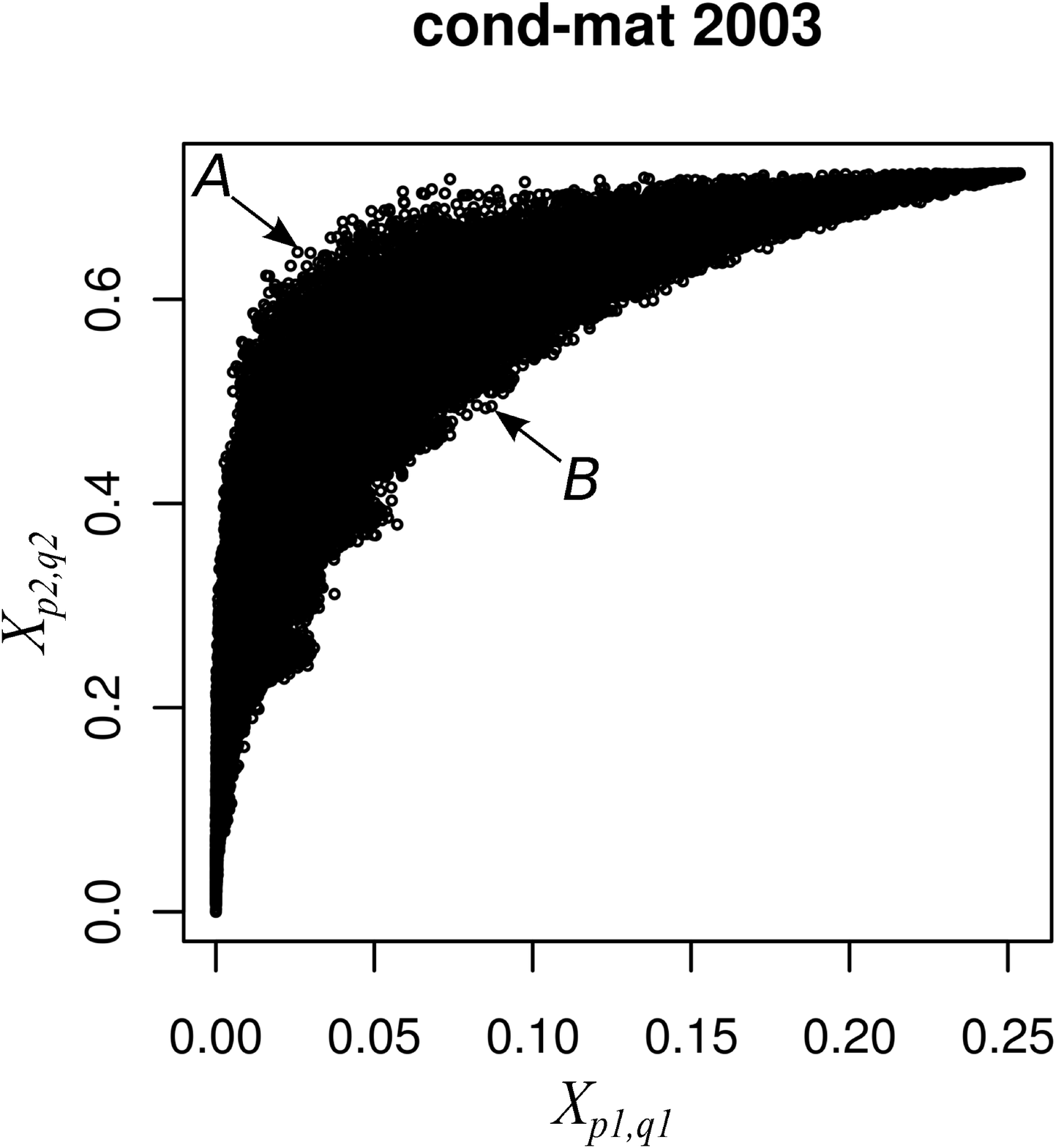}
\end{center}
\caption{
{\bf The mean number of infected nodes normalized to the total number of nodes
in the network in the regime $(p_1=0.1, q_1=0.9)$ (on the x-axis) 
versus the mean number of infected nodes normalized to the total number of nodes
in the network in the regime $(p_2=0.1, q_2=0.2)$ (on the y-axis) for the
cond-mat complex network and the same initially infected node. For a very 
large number of node pairs their relative epidemic ranking changes when the
first regime is changed 
to the second one. Points $A$ and $B$ marked in the plot give a clear example of
such a pair.} }
\label{fig:twopq} 
\end{figure}

The testing of the hypothesis formulated in the preceding paragraph on empirical
complex networks is carried out in the following way. Two pairs of SIR model
parameters $(p_1,q_1)$ and $(p_2,q_2)$ are selected. For each node of the
empirical complex network and for each pair of the parameter values, the average
number of infected nodes (i.e. the final size of the epidemic) for the disease
starting at that very node is calculated. In particular for a node $i$ the
quantities $X^i_{p_1,q_1}$ and $X^i_{p_2,q_2}$, the average numbers of infected
nodes normalized to the total number of nodes in the network for parameters
$(p_1,q_1)$ and $(p_2,q_2)$, respectively, are obtained. A plot with
$X^i_{p_1,q_1}$ and $X^i_{p_2,q_2}$ on the axes is constructed. For each
initially infected node a point is entered into the plot. An example of such a
plot obtained for the cond-mat network is presented in Fig. \ref{fig:twopq}.

The scattering of the points in the plot presented in Fig. \ref{fig:twopq}
vividly demonstrates the dependence of the epidemic ranking on the disease
spreading regime. In this plot two points $A$ and $B$ have been singled out to
show how the ranking of two nodes is altered if the dynamical regime of disease
spreading is changed. If there was no dependence of the epidemic ranking on 
parameters of the disease spreading model, the points in Fig. \ref{fig:twopq}
would be ordered in a monotonously growing curve. 
Analogous plots demonstrating the dependence of
the ranking on the dynamical regime have also been obtained for other studied empirical complex networks
and other combinations of $(p,q)$ pairs.

The results of simulations on empirical complex networks presented above clearly
illustrate the dependence of the node's ranking according to its
epidemic impact on the dynamical regime of spreading. There are many conceivable mechanisms how this dependence might
be realized in practice. In the remainder of this section we focus on an
analytically tractable example of the disease spreading where we demonstrate one
possible mechanism of dependence on the dynamical regime of spreading.

\begin{figure}[!ht]
\begin{center}
\includegraphics[width=0.6\textwidth]{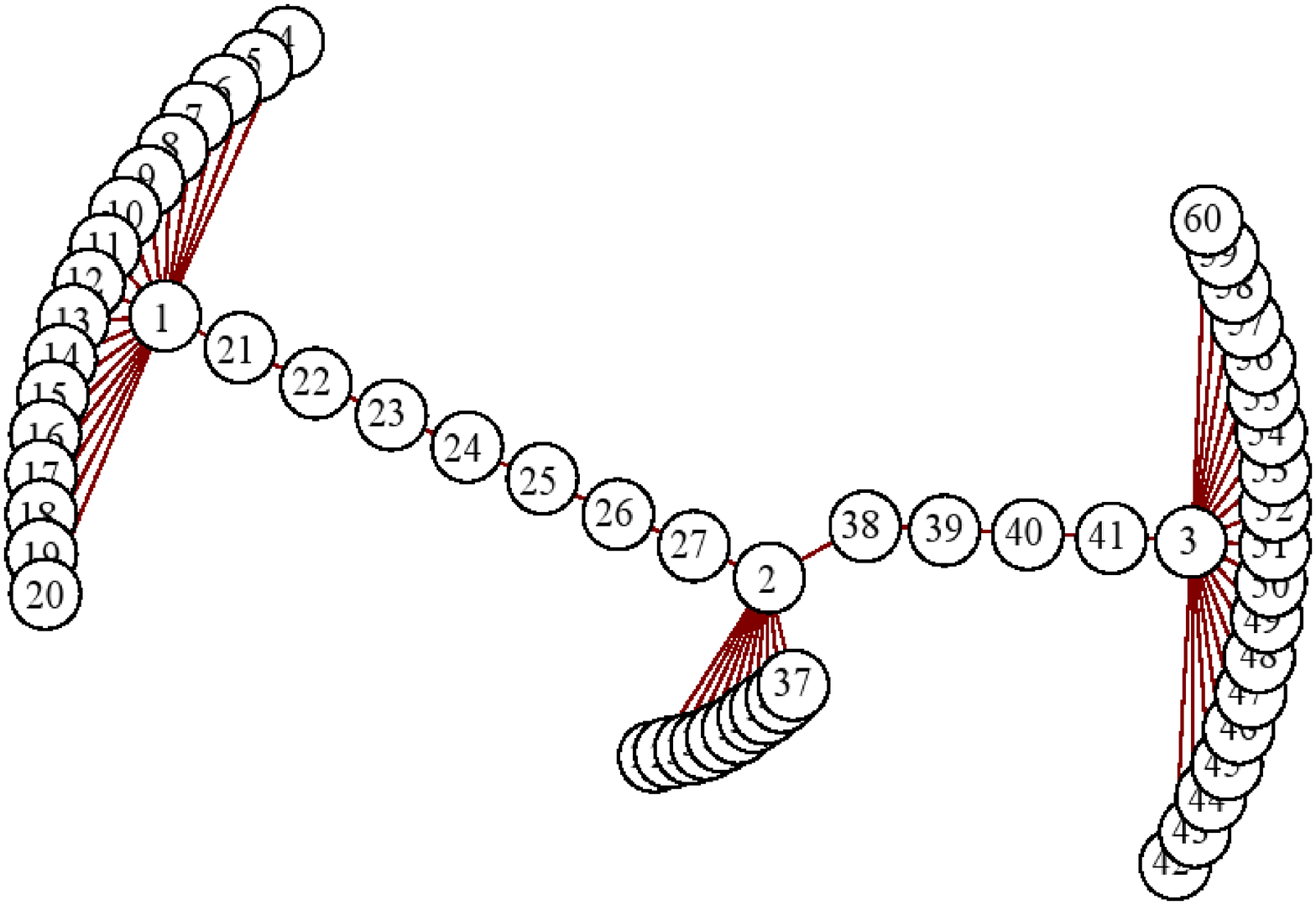}
\end{center}
\caption{ 
{\bf The artificial network with three distinguished nodes 1, 2, and 3 with
parameters $k_1=18$, $k_2=12$, $k_3=20$, $n_1=7$ and $n_3=4$ as defined in the
text (Online version in colour).} }
\label{fig:synthnet}
\end{figure}

We consider an artificial undirected network dominated by three nodes with high
degrees. Let us denote these nodes by $1$, $2$ and $3$ and let their degrees be
$k_1$, $k_2$ and $k_3$, respectively. One of these nodes, the node $2$, has a
central position in the network. It is connected to nodes $1$ and $3$ with
chains of length $n_1$ and $n_3$ respectively. The nodes connected to one of the
nodes $1$, $2$ and $3$ are said to belong to their respective stars. We are
interested in the situation where the central node $2$ has smaller degree than
the nodes $1$ and $3$, i.e. $k_2 < k_1$ and $k_2 < k_3$. The lengths of the
chains are comparable, i.e. $n_1 \simeq n_2$. An example of such a network with
$k_1=18$, $k_2=12$, $k_3=20$, $n_1=7$ and $n_3=4$ is presented in Fig.
\ref{fig:synthnet}.

The mechanism of disease spreading in this network is relatively simple. Since
the network has a tree topology, for an arbitrary initially infected node there
is a unique path in the network via which any other node can be infected. For a
node to get infected, its neighbor on the path connecting the studied node with
the initially infected node must be infected too. A formalism for the full
analytical description of the disease spreading in tree-like networks has been
recently developed in \cite{miPhysA}. In general, we consider a bipartite graph
with two classes of nodes. The class I contains $s$ nodes which are all in the
infected (I) state. The class II consists of $n$ nodes which are in the
susceptible (S) state. Every node from the class I is connected to all nodes
from the class II. The probability that the random variable $X_n^{(s)}$,
numbering eventually infected nodes in class II, acquires the value $k$ is
\cite{miPhysA}
\begin{eqnarray}
\label{eq:distrX}
&p^{(s)}_{n,k}& \equiv P(X_n^{(s)}=k) = \nonumber \\&=& q^s \binom{n}{k}
\sum_{l=0}^k \binom{k}{l} (-1)^l
\left(\frac{(1-p)^{n-k+l}}{1-(1-q)(1-p)^{n-k+l}}\right)^s \, .
\end{eqnarray}
Using the expression (\ref{eq:distrX}), we consider the expected number of
infected nodes for two different initially infected nodes. For the scenario in
which the spreading of the disease starts from the node $1$ the random variable
of the number of infected nodes is denoted by $Y_1$. For the second scenario in
which the spreading begins from the node $2$, the respective random variable is
denoted by $Y_2$. For the calculation of expected values of $Y_1$ and $Y_2$ we
need a particular instance of (\ref{eq:distrX}), namely $P(X_1^{(1)}=1)$. For a
selected initially infected node, in the studied network every node can be
infected only from one of its neighbors. Furthermore the process of disease
transfer from the infected to the susceptible node is an independent process for
all pairs of neighboring nodes consisting of one infected and one susceptible
node. The existence of the chain between the nodes $1$ and $2$ and the chain
between the nodes $2$ and $3$ is the most artificial element of the studied
network. It is needed to produce a large illustrative variation of the
expectation of $Y_1 - Y_2$. For shorter chains this variation would be smaller,
but still present.

The expected value of the number of infected nodes in the first scenario of
interest is given with the following expression:
\begin{eqnarray}
 \label{eq:expy1}
E(Y_1)&=& 1 + k_1 P(X_1^{(1)}=1)+ \sum_{i=2}^{n_1+1} (P(X_1^{(1)}=1))^i
\nonumber \\
&+& (k_2-1) (P(X_1^{(1)}=1))^{n_1+2} \nonumber \\ 
&+& \sum_{i=n_1+3}^{n_1+n_3+2}  (P(X_1^{(1)}=1))^i \nonumber \\
&+& (k_3-1) (P(X_1^{(1)}=1))^{n_1+n_3+3} \, .
\end{eqnarray}
Here the first term represents the initially infected node and the second term
represents the expected number of infected nodes in the star of the node $1$,
whereas the third term with the sum gives the expected number of the infected
nodes in the rest of the chain between nodes $1$ and $2$, including the node
$2$. The fourth term depicts the number of infected nodes in the rest of the
star of the node $2$, the fifth term with the sum represents the number of
infected nodes in the rest of the chain between the nodes $2$ and $3$, including
the node $3$, and the final sixth term gives the number of infected nodes in the
rest of the star of the node $3$.
      
\begin{figure}[!ht]
\begin{center}
\includegraphics[width=0.6\textwidth]{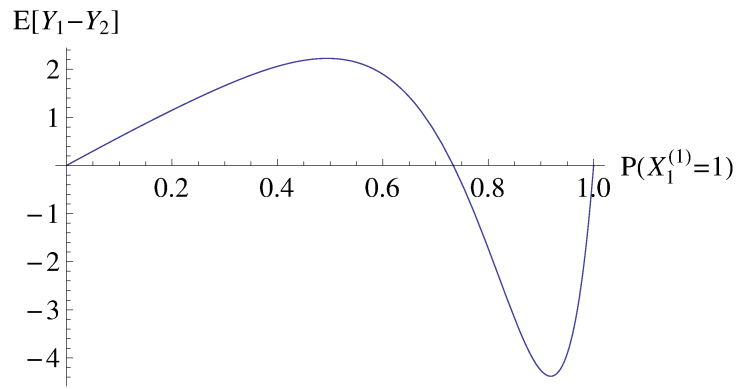}
\end{center}
\caption{
{\bf The difference of the expected numbers of infected nodes for scenarios
where the node 1 is the initially infected node $(Y_1)$ and the node 2 is the
initially infected node $(Y_2)$ (Online version in colour).}} 
\label{fig:diffY}
\end{figure}

The expected value of infected nodes in the second scenario is:
\begin{eqnarray}
 \label{eq:expy2}
E(Y_2)&=&1 + k_2 P(X_1^{(1)}=1) + \sum_{i=2}^{n_1+1} (P(X_1^{(1)}=1))^i
\nonumber \\
&+& (k_1-1) (P(X_1^{(1)}=1))^{n_1+2} + \sum_{i=2}^{n_3+1} (P(X_1^{(1)}=1))^i
\nonumber \\
&+& (k_3-1) (P(X_1^{(1)}=1))^{n_3+2} \, , 
\end{eqnarray} 
with respective terms as defined in (\ref{eq:expy1}).
A prominent feature of both expressions (\ref{eq:expy1}) and (\ref{eq:expy2}) is
that their dependence on $p$ and $q$ is captured by a single variable
$P(X_1^{(1)}=1)$. The difference of expectations of $Y_1$ and $Y_2$ as a
function of $P(X_1^{(1)}=1)$ is presented in Fig. \ref{fig:diffY}.

For small values of $P(X_1^{(1)}=1)$ the expectation of $Y_1$ dominates the
expectation of $Y_2$, but at larger values of $P(X_1^{(1)}=1)$ the expected
value of $Y_2$ is larger that the expected value of $Y_1$. For
$P(X_1^{(1)}=1)=0$ and $P(X_1^{(1)}=1)=1$ the expectations of these variables
are equal.

Fig. \ref{fig:diffY} shows that for $P(X_1^{(1)}=1)=0$ there is no spreading of
the disease. For small values of $P(X_1^{(1)}=1)$ the spreading of the disease
is contained and limited to the nearest neighbors. 
As $k_1 > k_2$, the expectation of $Y_1-Y_2$ is positive. As $P(X_1^{(1)}=1)$
grows, the disease progresses along the chains and for a sufficiently large
$P(X_1^{(1)}=1)$  the disease reaches the end of the chains. For a scenario with
node $1$ as the origin of the disease, in the regime of large $P(X_1^{(1)}=1)$
the disease spreads to the node $2$ and its star. However, the disease does not
spread to the node $3$ and its star. On the other hand, when the disease
originates in node $2$, owing to the central position of that node the disease
spreads to both nodes $1$ and $3$ and their respective stars in the regime of
sufficiently large $P(X_1^{(1)}=1)$. That is why the expectation of $Y_1-Y_2$
becomes negative in this regime. Finally, for $P(X_1^{(1)}=1)=1$ the disease
spreads to the entire network in both scenarios and the expectations of $Y_1$
and $Y_2$ are equal to the total number of nodes in the network.     

The analysis of the disease spreading in the network given in Fig.
\ref{fig:synthnet} serves as an illustration how the dependence of the
node's epidemic ranking on the dynamical regime of spreading might be realized. It is reasonable to assume that this
mechanism is just one of a broad class of mechanisms leading to dynamical
dependence of the epidemic ranking. Some of these mechanisms should also be
effective in networks with cycles.


The dependence of the epidemic ranking of a node on the disease spreading
dynamical regime has important implications in situations where some preparative
action needs to be taken before the dynamical regime of disease spreading is
known. An example of such a situation is the design of security and public
health systems for countering the disease spreading. These protective system
should, at least to some extent, function for virtually all disease spreading
regimes.
Some sort of average epidemic ranking, with averaging taken over all disease
spreading regimes, becomes an essential ingredient for decisions on the
structure of protective systems and efficient mitigation strategies. On the
other hand, knowing some average impact of a node over all disease spreading
regimes is important in its own right as a measure of epidemic importance of a
node in the network. 
A useful framework for the calculation of the needed averages, the {\em phase
diagram of epidemic spreading}, is described in the following subsection.  The
averaging procedures which take into account the dependence of the
ranking on the dynamical regime of spreading are discussed in the final subsection of this section.

%

\subsection*{Phase diagram of epidemic spreading}
\label{phase}

%
For a fixed underlying complex network and a fixed initially infected node the
outcome of the disease spreading still strongly depends on the properties of the
disease itself, measured by parameters $p$ and $q$ in the SIR epidemic model. A
very useful concept for the understanding and representation of the epidemic
impact in the studied complex network for different values of $p$ and $q$ and a
fixed initially infected node, named {\em the phase diagram of epidemic
spreading}, has been recently introduced in \cite{miPhysA}.

In the phase diagram of epidemic spreading we consider the parametric space of
the SIR model which is a $[0,1] \times [0,1]$ square. For each variable chosen
to measure the impact of disease spreading a phase diagram can be constructed.
Such variables useful in describing the impact of disease spreading are e.g. the
average number of infected nodes (i.e. the final size of the epidemic) or the
cumulative probability for a finite epidemic range \cite{miPhysA}. The phase
diagram of epidemic spreading is constructed in the following way: for each pair
of allowed $(p,q)$ parameters a value of the variable $X_{p,q}$ measuring the
extent of disease spreading is determined (analytically or in simulations) and
all triplets $(p,q,X_{p,q})$ are organized in a single diagram.

The phase diagram of epidemic spreading provides valuable insight into the
bimodal character of disease spreading, i.e. equilibrium between the local
containment of the disease and epidemic outbreak affecting the entire system
\cite{miPhysA}. Furthermore, it is a useful tool for searching for generic
properties of disease spreading across different complex networks and initially
infected nodes \cite{miPhysA}. Finally, the phase diagram of epidemic spreading
provides a global insight into a full set of disease spreading regimes. With a
phase diagram of epidemic spreading available, one just needs to decide which
averaging procedure is relevant and should be applied for the ranking
calculation.

\subsection*{Epidemic centrality - an epidemic impact measure}

\label{centr}

To take into account the dependence of the epidemic ranking on the disease
spreading dynamical regime, it is necessary to find a robust way to combine the
effects of the entire parametric space, as stated in the preceding subsection.
The concept of phase diagram of epidemic spreading  lends itself as a natural
framework for the definition of such a robust combination. Namely, a natural
candidate for the measure of epidemic impact is some weighted average of the
phase diagram of epidemic spreading over parametric space. We call this measure
of epidemic impact {\em epidemic centrality} and for the node $i$ we denote it
by $Z^i$.

The simplest option is the uniform weight function. All disease spreading
regimes are taken with equal weights in the calculation of the ranking (epidemic
centrality), allowing us to express the epidemic centrality $Z^{i}$ as an
integral over parametric space. In particular, for the SIR model one obtains
\begin{equation}
 \label{eq:uniform}
Z^{i}=\int_0^1 dp \int_0^1 dq \, X^{i}_{p,q} \, .
\end{equation}
The assumption of uniform weighing of all disease spreading regimes could be
contested as overly simplifying since a large majority of known contagious
diseases have comparable or at least not drastically different transmission
rates and average recovery times. Still, recent attempts of synthesis of
artificial microorganisms \cite{Venter} and nonspecific transmission patterns of
some diseases among amphibian populations \cite{amph} warn us that diseases with
nonstandard spreading regimes might pose significant risks in the future. These
observations support uniform weighting of all spreading regimes in the averaging
procedure of ranking calculation. In any case, (\ref{eq:uniform}) is a good
starting point which should provide a good approximation of the ranking.

In general, the averaging should be performed using some nonuniform weight
function $w(p,q)$. All available additional information on the epidemic risks
posing a threat should be incorporated into that function. For example, if it is
known that the spreading regimes of the diseases posing the greatest threat are
constrained to a segment of the parametric space, then the weight function should
have a peak in this part of the parametric space. The epidemic centrality is
then calculated as 
\begin{equation}
 \label{eq:general}
Z^{i}=\int_0^1 dp \int_0^1 dq  \, w(p,q) \, X^{i}_{p,q} \, .
\end{equation}
 
\begin{figure}[!ht]
\begin{center}
\includegraphics[width=0.8\textwidth]{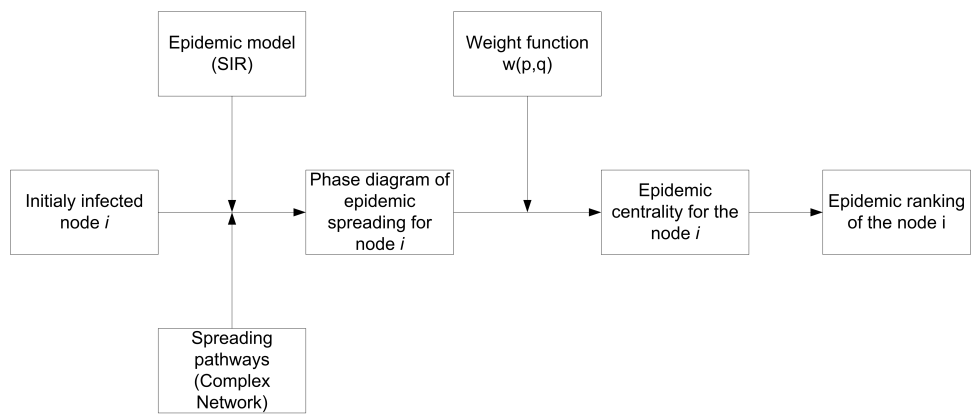}
\end{center}
\caption{
{\bf The schematic representation of the procedure for the calculation of
epidemic centrality.} }
\label{fig:flow}
\end{figure}

The procedure of calculation of epidemic centrality is schematically depicted in
Fig. \ref{fig:flow}. The idea of proper averaging over different disease spreading parameters has been recently also addressed in SIS model \cite{Youssef}.
In that paper, Youssef et al. argue that measures such as the epidemic threshold or the number of infected nodes in the asymptotic steady state, taken 
separately, do not give the appropriate measure of network robustness to epidemic. They introduce a useful novel measure, defined at the level of the entire network,
which they call {\em viral conductance (VC)}. Viral conductance is essentially an average of number of infected nodes in the asymptotic steady state for all
SIS model parameters above epidemic threshold. The similarity of the concepts of viral conductance for SIS model and epidemic centrality for SIR model lies in 
the premise that numbers of infected nodes for various disease spreading parameters need to be combined to obtain relevant measures of importance of network 
topologies (in the case of viral conductance) or node positions (in the case of epidemic centrality) in disease spreading. However, viral conductance and 
epidemic centrality are defined for SIS and SIR, respectively, which are substantially different models used for modeling of spreading of different diseases.
The concept of epidemic centrality is defined for {\em every} individual node in the network and it furthermore allows the ranking of the nodes according to 
their epidemic impact. The concept of viral conductance, on the other hand, is defined at the level of the entire network structure. 
Finally, the motivation for the introduction
of epidemic centrality is a much broader phenomenon of dependence of epidemic ranking on disease spreading parameters, discussed in detail in this paper. The concept of 
viral conductance is largely complementary to epidemic centrality, given their different properties and areas of applicability. 
Nevertheless, the proposal of Youssef et al. in SIS, along with the concept of epidemic centrality introduced in our paper, reasserts 
the need to evaluate the importance of
network structure in disease spreading for many disease spreading parameters simultaneously.

%

\section*{Results and Discussion}

\label{disc}

\begin{figure}[!ht]
\begin{center}
\includegraphics[width=1.0\textwidth]{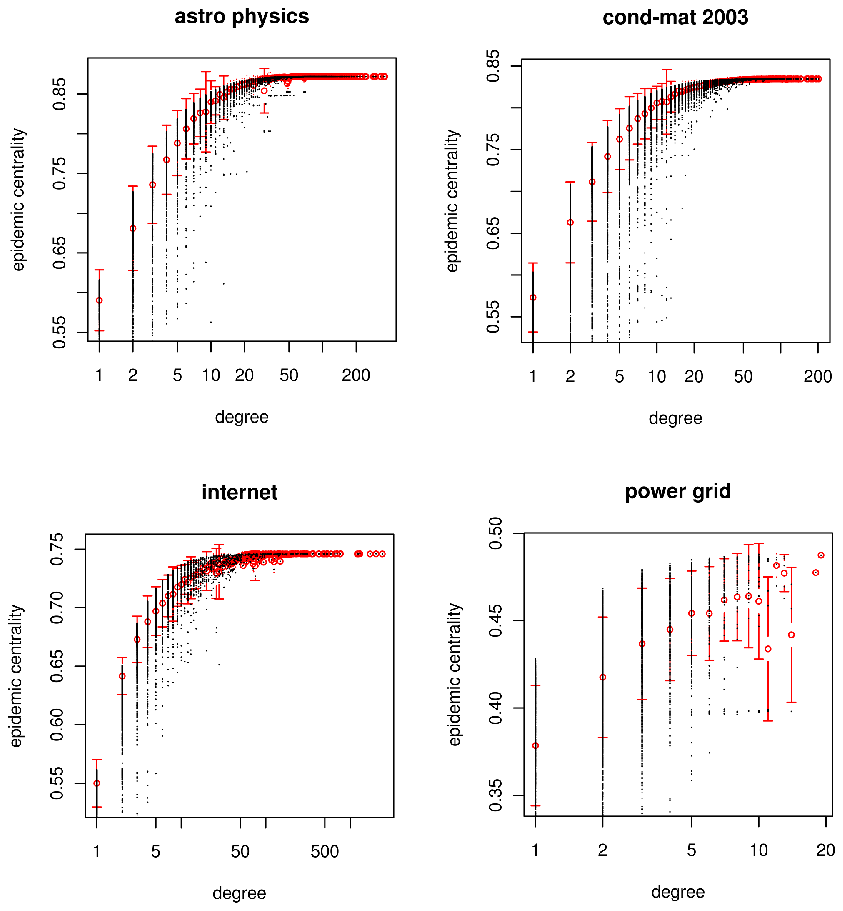}
\end{center}
\caption{
{\bf The epidemic centrality versus the degree of the initially infected node
for the studied empirical complex networks: astro-ph (top left), cond-mat 2003
(top right), internet (bottom left) and power grid network (bottom right). Black dots represent epidemic centrality values of individual 
nodes and red circles with error bars represent the average value and the standard 
deviation of epidemic centrality for a given degree.}}
\label{fig:degreeall}
\end{figure}

We first focus on results obtained using the uniform weight function
(\ref{eq:uniform}). Although our main discussion is focused on the comparison of
epidemic centrality with structural measures such as node degree, we would like to stress 
that epidemic centrality is a well defined and motivated measure of the epidemic impact on
its own. The said comparison does not serve as a validation of the epidemic centrality, but as 
a part of analysis on the importance of the nodes which are structurally peripheral and 
the nodes which are structurally central.
The dependence of epidemic centrality on the node degree is 
studied for four empirical complex networks (astro-ph, cond-mat 2003, internet, 
power grid) and their respective epidemic centralities are shown in Fig.
\ref{fig:degreeall}.
All studied networks share some common properties. The epidemic centrality in
general grows with the degree of the initially infected node although 
considerable scattering for the same degree exists. The average 
of epidemic centralities of all nodes with the same degree grows with the degree
of 
the initially infected node, whereas the standard deviation as a measure of
scattering 
decreases with the degree of the initially infected node, as depicted in Fig.
\ref{fig:degreeall}.
For astro-ph, cond-mat 2003 and internet networks the scattering reduces
considerably for high 
degree nodes and for power-grid network scattering is present even for the high
degrees.

\begin{figure}[!ht]
\begin{center}
\includegraphics[width=1.0\textwidth]{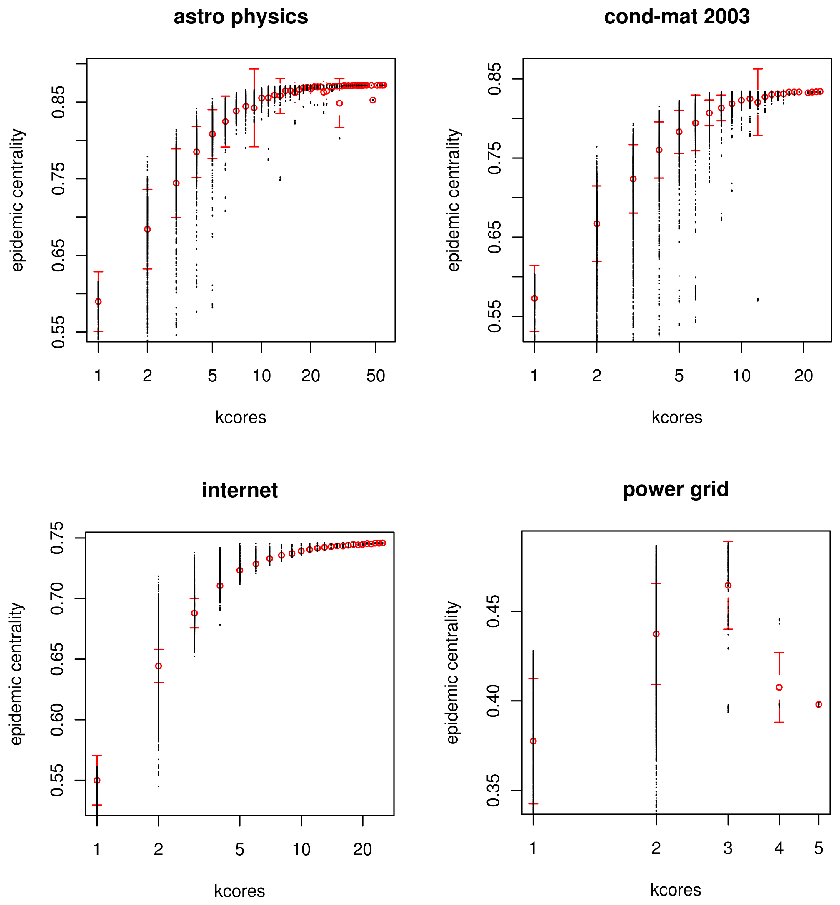}
\end{center}
\caption{ 
{\bf The epidemic centrality versus the k-cores of the initially infected node
for the studied empirical complex networks: astro-ph (top left), cond-mat 2003
(top right), internet (bottom left) and power grid network (bottom right).  Black dots represent epidemic centrality values of individual nodes and 
red circles with error bars represent the average value and the standard 
deviation of epidemic centrality for a given k-cores value.  
}}
\label{fig:kcoresall}
\end{figure}

The dependence of epidemic centrality of infected nodes on the k-cores variable
of each 
initially infected node for the four studied networks is presented in Fig.
\ref{fig:kcoresall}.
For astro-ph, cond-mat 2003 and internet networks, the epidemic centrality in
general grows with  k-cores of initially infected nodes. As depicted in 
Fig. \ref{fig:kcoresall},
the epidemic centrality of initially infected nodes 
with the same k-cores value grows with the k-cores, whereas the standard
deviation as a measure 
of scattering decreases with the k-cores value. The power grid network 
exhibits somewhat different behavior. Namely, the epidemic centrality of
initially infected nodes with 
the same k-cores value first increases and then decreases, but the standard
deviation decreases 
with the k-cores values. The interpretation of this peculiarity is obscured by a
very small 
range of k-cores values present in the power grid network.

\begin{figure}[!ht]
\begin{center}
\includegraphics[width=1.0\textwidth]{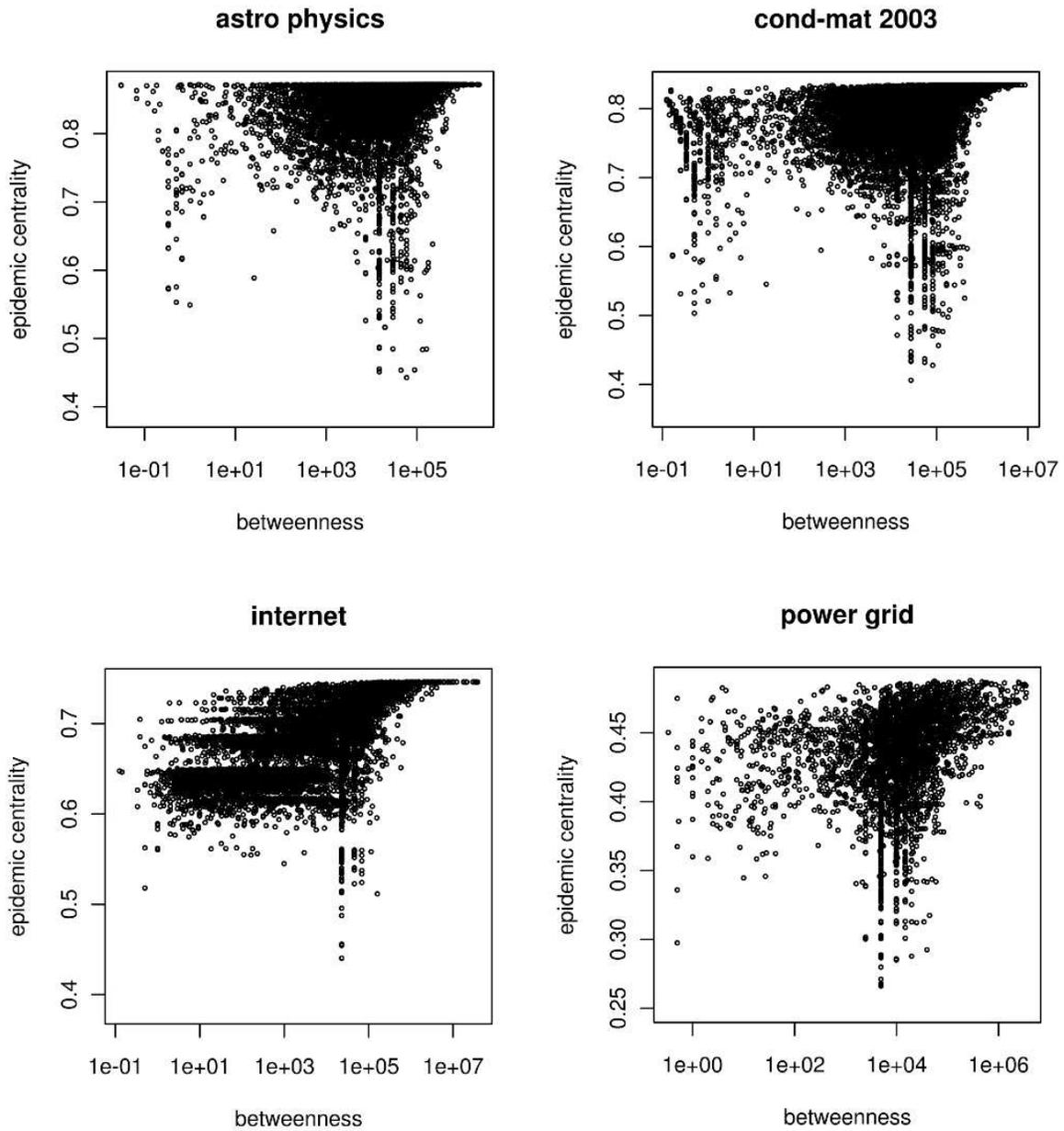}
\end{center}
\caption{
{\bf The epidemic centrality versus the betweenness of the initially infected
node for the studied empirical complex networks: astro-ph (top left), cond-mat
2003 (top right), internet (bottom left) and power grid network (bottom right). 
  }}
\label{fig:betall}
\end{figure}

The dependence of epidemic centrality on betweenness of the initially 
infected nodes for four studied networks is displayed in Fig. \ref{fig:betall}.
From all these four figures it is evident that there is no clear relation
between the epidemic centrality and betweenness of the initially infected node.
Although no clear relation between 
the epidemic centrality and the betweenness of the initailly infected nodes
can be 
identified, it is striking that the patterns in all plots in Fig.
\ref{fig:betall} 
are very similar 
in general and exhibit very similar peculiar details. In particular, in all
studied networks there are two intervals of betweenness where epidemic
centrality has dispersion larger than average: the first at small values of
betweenness and the second close to the largest values of betweenness for the
studied network.

\begin{figure}[!ht]
\begin{center}
\includegraphics[width=1.0\textwidth]{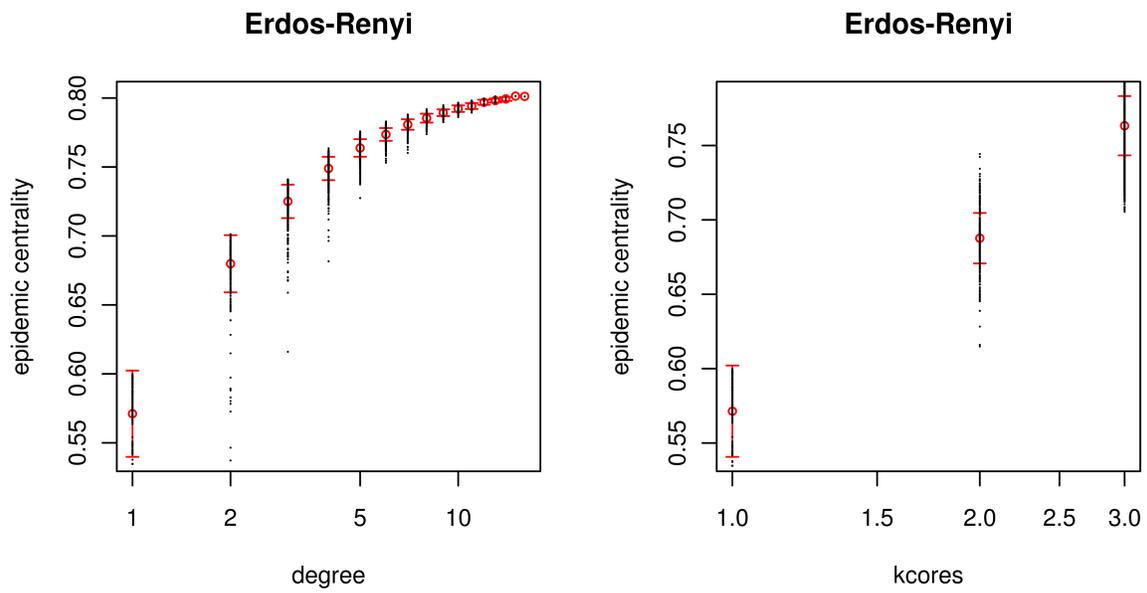}
\end{center}
\caption{
{\bf The epidemic centrality versus the degree (left) and the k-cores (right) of the initially infected
node for an Erdos-Renyi network G(5000, 0.001). 
  }}
\label{fig:ER}
\end{figure}

Finally, the dependence of epidemic centrality on degree and k-cores value of the 
initially infected node for the simulated Erdos-Renyi networks is presented in Fig.
\ref{fig:ER}. This figure clearly shows a qualitatively same dependence of 
epidemic centrality on the studied structural variables. This finding indicates that the 
observed properties of epidemic centrality are not restricted to the studied empirical complex
networks or networks with broad degree distributions. Establishing a possible universality 
of some epidemic centrality properties over broad classes of networks would, however, require 
a far more extensive study which is beyond the scope of this paper.  

Although the relation between epidemic centrality on the one hand and structural
variables such as degree and k-cores on the other hand is clearly nonlinear, it
is also important to know how similar are rankings based on degree or k-cores
value to the epidemic ranking based on epidemic centrality. To establish this
similarity we relabel nodes according epidemic ranking (the highest ranked node
is relabeled $1$ and the $i^{th}$ ranked node is relabeled as $i$) and then
produce the sequence of rankings according to degree (or k-cores value) and to
each node $i$ we assign the structural ranking $r_{\mathrm{struct}}(i)$. Then we
calculate Spearman's rank correlation coefficient of sequences $i$ and
$r_{\mathrm{struct}}(i)$. The results for the studied networks are presented in
Table \ref{tab:rankrank}. The obtained results show that for astro-ph, cond-mat
2003 and internet networks ranking based on degree or k-cores value is very
similar to epidemic ranking. For the power grid network the correlation is
notably lower, but still considerable.  

The most interesting feature observed in all networks discussed in this section
is a very small ratio of the largest and the smallest epidemic centrality in the
network. The epidemic centrality of nodes that are completely peripheral in the
structure of the network (either in terms of their degree of k-cores value) is
only about a factor 2 smaller than the epidemic centrality of the nodes that we
would describe as central from the structural point of view. Given that the
variation in degree in all studied networks (except power grid) goes up to
several orders of magnitude and that k-cores variable acquires values of more
than 20, it is intriguing that the epidemic centrality is so insensitive to this
variation. 

A reasonable question is how much the findings of the preceding paragraph depend
on the choice of the weight function. In particular, it would be important to
learn if the relative insensitivity of epidemic centrality on structural
variables is a consequence of the uniform weight function, see
(\ref{eq:uniform}). As there is an  unlimited number of  choices for the
nonuniform weight function, we restrict ourselves to beta distribution
$f_{\alpha,\beta}(x)=\frac{\Gamma(\alpha+\beta)}{\Gamma(\alpha) \Gamma(\beta)}
x^{\alpha-1} (1-x)^{\beta-1}$  \cite{wikibeta} as a sufficiently broad class
able to approximate well any weight function that might be of interest.

\begin{figure}[!ht]
\begin{center}
\includegraphics[width=1.0\textwidth]{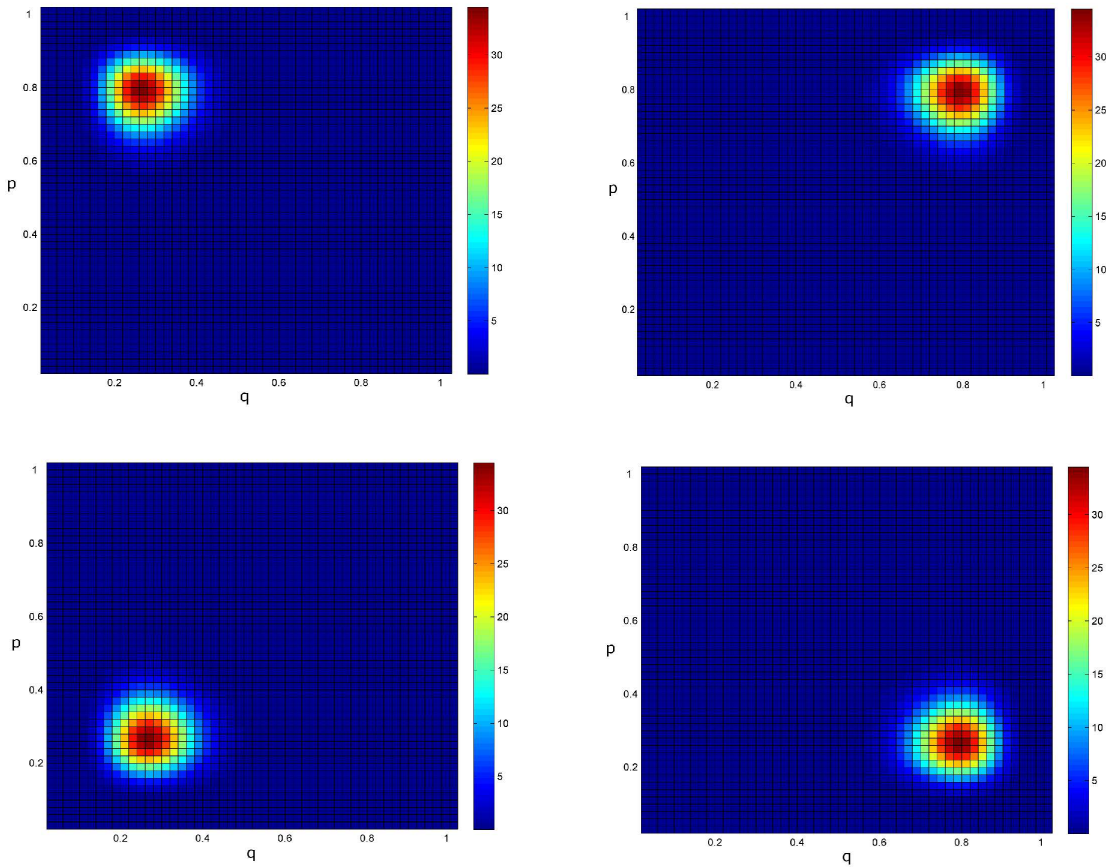}
\end{center}
\caption{ 
{\bf Contour plots of four weight functions localized in the region of small $q$ and large $p$ ($w_A(p,q)$, top left), large $q$ and large $p$ 
($w_B(p,q)$, top right),
small $q$ and small $p$ ($w_C(p,q)$, bottom left) and large $q$ and small $p$ ($w_D(p,q)$, bottom right). For the definitions of $w_{A,B,C,D}(p,q)$ see the section 
Results and Discussion.} }
\label{fig:fourweights}
\end{figure}
\begin{figure}[!ht]
\begin{center}
\includegraphics[width=1.0\textwidth]{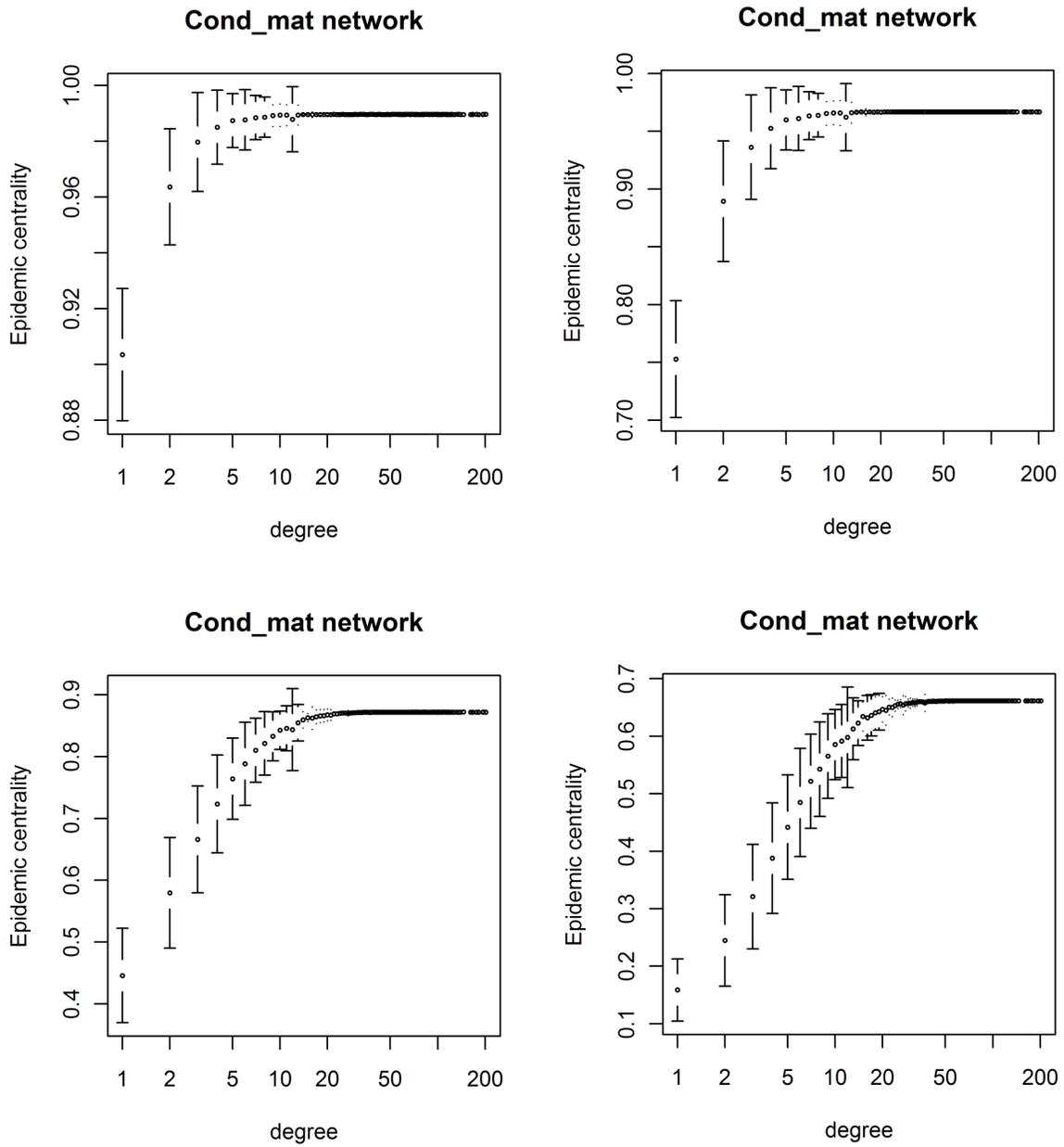}
\end{center}
\caption{ 
{\bf The average values and standard deviations of epidemic centrality versus node degree for cond-mat 2003 network for four different weight functions depicted in
Fig. \ref{fig:fourweights}.  } }
\label{fig:avepicenfourweights}
\end{figure}

We find that the qualitative conclusions presented for the uniform $w(p,q)$
hold also for other weight functions localized in specific parts of the parametric
space. To demonstrate this, we choose four  
weight functions $w(p,q)$ which are well localized in four quadrants of the $(p,q)$
parametric space. We select the weight functions of the form 
$w(p,q;\alpha_1 ,\beta_1,\alpha_2 ,\beta_2)= f_{\alpha_1,\beta_1}(p) \cdot f_{\alpha_2 ,\beta_2}(q)$, where $f_{\alpha,\beta}(x)$ is the beta distribution.
Particular forms of the functions are
$w_A(p,q) = w(p,q;\alpha_1 = 30 ,\beta_1 = 10,\alpha_2 = 10 ,\beta_2 = 30)$, 
$w_B(p,q) = w(p,q;\alpha_1 = 30 ,\beta_1 = 10,\alpha_2 = 30 ,\beta_2 = 10)$, 
$w_C(p,q) = w(p,q;\alpha_1 = 10 ,\beta_1 = 30,\alpha_2 = 10 ,\beta_2 = 30)$, and 
$w_D(p,q) = w(p,q;\alpha_1 = 10 ,\beta_1 = 30,\alpha_2 = 30 ,\beta_2 = 10)$,  
which are presented in Fig. \ref{fig:fourweights}. For the cond-mat 2003 network, the relation of the 
node degree and the average epidemic centrality for that node degree is given in Fig.
\ref{fig:avepicenfourweights} for the four choices for the weight function. Although the range of values of average 
epidemic centrality varies with the chosen weight function, the average epidemic centralities 
remain of the same order of magnitude even as the degree varies several orders of magnitude.
An equivalent conclusion can be drawn on the relation of the k-cores value and the average
epidemic centrality for a particular k-cores value. Furthermore, equivalent conclusions can be
reached for astro-ph, internet and power networks. It is important to observe that for the small 
values of $p$ and large values of $q$ (in particular for the weight function $w_{D}(p,q)$) the values 
of epidemic centrality for some nodes of small degree may become rather small (much smaller than the average 
epidemic centrality for a given degree). However, the average epidemic centrality (and a majority 
of epidemic centrality values of individual nodes) remains of the same order 
of magnitude as for the large degrees. 









\begin{figure}[!ht]
\begin{center}
\includegraphics[width=1.0\textwidth]{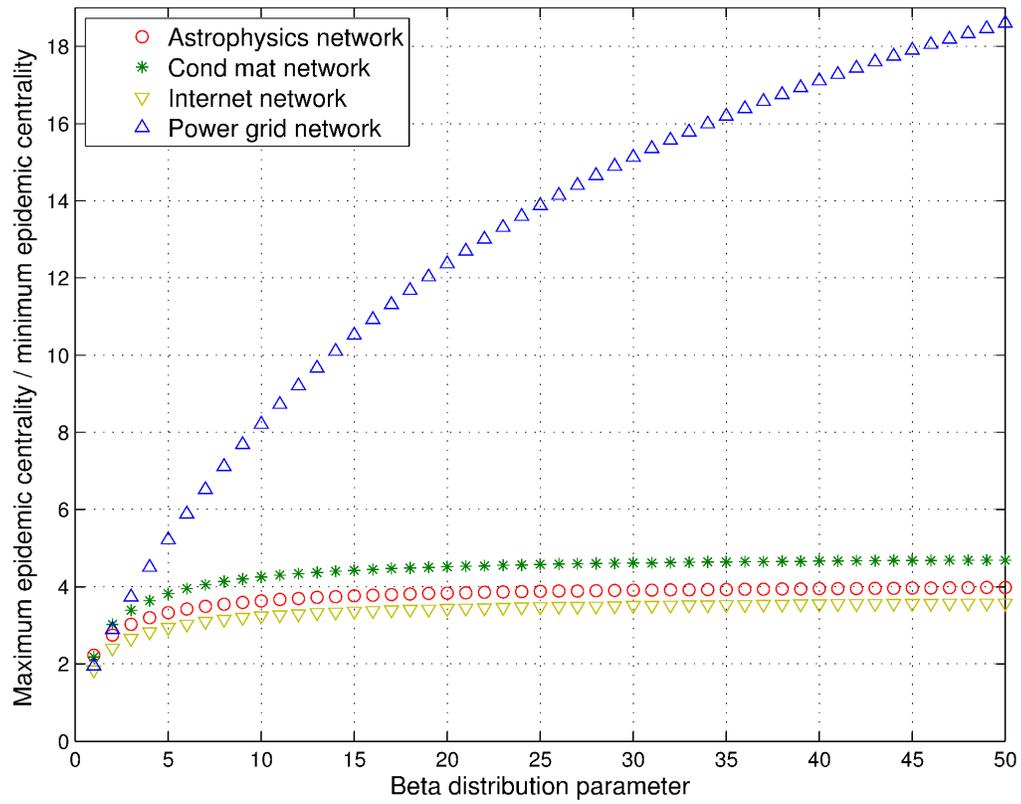}
\end{center}
\caption{ 
{\bf The ratio of maximal to minimal average epidemic centrality of nodes with the same degree for four studied networks
for different localizations of the weight function measured by the parameter
$\alpha$ (Online version in colour).} }
\label{fig:ratioall}
\end{figure}

To further elaborate on the dependence of our results on the choice of weight
function
and the problem of similarity/distinction of uniform and nonuniform weight
functions, we consider the following analysis. We consider a family of weight
functions $w(p,q)=f_{\alpha,\alpha}(p) f_{\alpha,\alpha}(q)$ that contains a
uniform weight function as a special case for $\alpha=1$ and
$f_{\alpha,\alpha}(x)$ is symmetrically centered around its expectation value
$x=1/2$. Starting from $\alpha=1$ and increasing the value of $\alpha$ we go
from the uniform weight functions to the more and more localized ones. The value
of $\alpha$ serves as a measure of localization of the weight function. For each
of $\alpha$ values we calculate the ratio of maximal and minimal average
epidemic centrality for nodes with the same degree and plot this ratio as a
function of $\alpha$. These plots for all studied networks are presented in Fig.
\ref{fig:ratioall}. For astro-ph, cond-mat 2003 and internet networks even for
very localized weight functions the studied ratio remains very close to its
value for the uniform weight function. In the case of power grid network, the
ratio grows with localization, although at a decreasing rate. As already
observed at other places in this paper, the power grid exhibits different
behavior than other studied complex networks. A possible interpretation of this 
difference might lie in the fact that the degree distribution for the power grid
network is exponential, whereas other empirical complex networks have a 
broad degree distribution.

The studies presented so far point to a somewhat surprising general conclusion
with very important practical consequences. The epidemic centrality as an
average measure of epidemic impact is much more homogeneous among nodes of
studied networks than are its structural properties. In the studied networks, the epidemic centrality of a node does show growing trends with its degree and k-cores value, but with a
rather low sensitivity to these structural variables. For example, whereas the
degree of nodes may differ several orders of magnitude, the epidemic centrality
changes within the factor of a few. A natural conclusion imposing itself is
that, as far as the epidemic centrality could be taken as a measure of the 
epidemic impact, the importance of structurally
peripheral nodes in the studied networks should be much higher than their structural variables might imply. 
Although our conclusions on the importance of the structurally peripheral nodes can be strictly applied only to networks and 
weight functions studied in this paper, a diverse character of studied complex networks (empirical vs. synthetic, ER vs. networks with 
broad degree distributions) and weight functions (uniform vs. localized), indicates that underestimated epidemic 
impact of structurally peripheral nodes might be a phenomenon valid in epidemic spreading in the SIR model in a much 
broader class of complex networks.  


The problem of allocation of resources dedicated to countering the disease
spreading (see e.g. \cite{Forster,Rowthorn}) should be definitely strongly
influenced by our findings on epidemic centrality. If we adopt a strategy that
the amount of resources allocated to a certain node should be proportional to
its epidemic centrality, then a logical conclusion is that virtually all nodes
in the network should receive comparable amount of resources. Another
implication is that allocating a lot of resources to hubs and nodes with high
k-cores values does not necessarily make the entire network more resilient to
epidemic outbreaks. To the contrary, the allocation of large amounts of
resources to structurally central nodes necessarily leaves structurally
peripheral nodes without resources although they are associated to a comparable
epidemic impact.

The findings of the preceding paragraph rest on a simple assumption that the
disease spreading can be contained only at its source. i.e. the initially
infected node. In that case it is necessary to allocate as much resources to the
node as the epidemic impact would be if the disease escaped the initially
infected node. This assumption is certainly only approximately true, but in our
opinion it captures the leading contribution to the risk associated with the
epidemic. Moreover, it is especially applicable to the class of situations of a
newly introduced and rapidly spreading pathogen when standard resources such as
vaccines and medicines are not available. 

The finding that the nodes having very different structural variables actually
have comparable epidemic centralities does not imply that the roles of these
nodes in the process of disease spreading are the same. The fact that two nodes
$N$ and $M$ have comparable epidemic centralities just means that the mean
number of infected nodes, appropriately averaged over the parametric space, will
be comparable if the disease spreading starts at the node $M$ or the node $N$.
In the actual process of spreading, hubs have a far more prominent role than the
peripheral nodes. If the disease spreading starts at a peripheral node, the
spreading is slow until the disease reaches some of the hubs and then the
spreading (measured by the number of infected nodes) accelerates. 
The epidemic centrality describes the final outcome of the epidemic and not its
precise temporal development.

Finally, entire discussion in this section was focused on the implications of
the introduced concept of epidemic centrality in epidemiology. Our findings also
apply to e.g. problems of spreading of ideas and trends in social networks. The
relative insensitivity of epidemic centrality (or its counterpart in spreading
of ideas and trends in social networks) to structural measures of centrality
such as node degree or k-cores value might play an important role in
understanding of social dynamics on these networks. The possibility that the
capacity of spreading of new ideas or imposing new trends might not be an
exclusive privilege of highly connected nodes deserves further elaboration.

In conclusion, the structure of spreading pathways and the dynamics of disease transmission are
intertwined in a very complex way. Any ranking of nodes according to epidemic
impact based exclusively on structural arguments is therefore inadequate. The
first principal result of this paper is that the epidemic ranking depends
crucially on the disease spreading regime. If we are interested in finding some
ranking of initially infected nodes that does not depend on specific spreading
regime, the entire parametric space has to be taken into account using
appropriate averaging procedures. 
We introduce
epidemic centrality as a measure of epidemic centrality based on averaging over
entire parametric space. 
The second major
result of this paper is that the variation of epidemic centrality of initially
infected nodes is much smaller than the variation of their degrees or k-cores
values. This finding indicates that the epidemic risk associated to the structurally
peripheral nodes might be much larger than their degrees or k-cores values would
imply. If the spreading of the disease can be stopped only at its source, the
optimal distribution of resources dedicated to stopping the spreading should be
proportional to epidemic centralities of the initially infected nodes. 
The concept of epidemic
centrality merits further elaboration and its extension to other epidemiological
models and more realistic complex networks. These tasks, together with the
application of these results beyond epidemiology represent short term goals of
future research. 

\section*{Acknowledgments}
The authors would like to thank Petra Klepac for valuable comments on the
manuscript and useful pointers to the literature. The work of M. \v{S}. is
financed by Biomedical Research Council of A*STAR, Singapore and by Ministry of
Education Science and Sports of the Republic of Croatia under Contract No. 
036-0362214-1987 and 098-1191344-2860. The work of H. \v{S}. is supported by the
Ministry of Education Science and Sports of the Republic of Croatia under
Contract No. 098-0352828-2863.


\bibliography{epirisk}

\section*{Figures}

%

\FloatBarrier

\section*{Tables}

\begin{table}[!ht]
\caption{ 
\bf{Spearman's rank correlation coefficient of epidemic ranking and ranking
based on node degree or k-cores value for four studied complex networks.}}
\begin{tabular}{|c|c|c|}
\hline
& degree & k-cores \\
\hline
astro-ph & 0.97 & 0.96  \\
\hline
cond-mat & 0.94 & 0.93  \\
\hline
internet & 0.91 & 0.92 \\
\hline
power grid & 0.66 & 0.68 \\
\hline
\end{tabular}
\label{tab:rankrank}
\end{table}

\end{document}